\documentclass[12pt,fleqn]{article}
\usepackage{amsmath}

\begin{document}

\title{\textbf{On bosonic limits of two recent supersymmetric
extensions of the Harry Dym hierarchy}}

\author{\textsc{S.~Yu.~Sakovich\footnote{Home institution:
Institute of Physics, National Academy of Sciences, 220072
Minsk, Republic of Belarus. Contact e-mail: saks@pisem.net}}%
\bigskip\\{\footnotesize Mathematical Institute, Silesian
University, 74601 Opava, Czech Republic}}

\date{}

\maketitle

\begin{abstract}
Two generalized Harry Dym equations, recently found by Brunelli,
Das and Popowicz in the bosonic limit of new supersymmetric
extensions of the Harry Dym hierarchy [J.~Math.\ Phys.\
44:4756--4767 (2003)], are transformed into previously known
integrable systems: one---into a pair of decoupled KdV equations,
the other one---into a pair of coupled mKdV equations from a
bi-Hamiltonian hierarchy of Kupershmidt.
\end{abstract}

\section{Introduction}

Integrable supersymmetric differential equations have been
attracting much attention in modern mathematical physics and
soliton theory (see, e.g., \cite{KK} and references therein).
Supersymmetric extensions of known integrable bosonic (or
classical) systems are of particular interest, because, if the
number $N$ of Grassmann variables is greater than one, those
extensions can generate, in their bosonic limits, some new
integrable classical systems which generalize the initial ones.

Recently, Brunelli, Das and Popowicz \cite{BDP} studied
supersymmetric extensions of the Harry Dym hierarchy, and found,
as bosonic limits of $N = 2$ supersymmetric extensions, the
following two new classical generalizations of the Harry Dym
equation:
\begin{equation}
\label{e1}
\begin{split}
w_{0,t} & = \tfrac{1}{2}
\left(
w_0^{-1/2}
\right)_{xxx} , \\
w_{1,t} & = \tfrac{1}{64}
\left(
-16 w_{1,xxx} w_0^{-3/2}
+ 96 w_{1,xx} w_{0,x} w_0^{-5/2}
\right. \\
& + 72 w_{1,x} w_{0,xx} w_0^{-5/2}
-258 w_{1,x} w_{0,x}^2 w_0^{-7/2}
- 6 w_{1,x} w_1^2 w_0^{-7/2} \\
& \left.
+ 9 w_1^3 w_{0,x} w_0^{-9/2}
-108 w_1 w_{0,xx} w_{0,x} w_0^{-7/2}
+ 219 w_1 w_{0,x}^3 w_0^{-9/2}
\right) ,
\end{split}
\end{equation}
and
\begin{equation}
\label{e2}
\begin{split}
w_{0,t} & = \tfrac{1}{16}
\left(
8 \bigl( w_0^{-1/2} \bigr)_{xxx}
- 6 w_{1,x} w_1 w_0^{-5/2}
+ 9 w_1^2 w_{0,x} w_0^{-7/2}
\right) ,\\
w_{1,t} & = \tfrac{1}{32}
\left(
- 8 w_{1,xxx} w_0^{-3/2}
+ 48 \left( w_{1,x} w_{0,x} \right)_{x} w_0^{-5/2}
\right. \\
& - 144 w_{1,x} w_{0,x}^2 w_0^{-7/2}
- 6 w_{1,x} w_1^2 w_0^{-7/2} \\
& + 9 w_1^3 w_{0,x} w_0^{-9/2}
+ 12 w_1 w_{0,xxx} w_0^{-5/2} \\
& \left.
-126 w_1 w_{0,xx} w_{0,x} w_0^{-7/2}
+ 177 w_{0,x}^3 w_1 w_0^{-9/2}
\right) ,
\end{split}
\end{equation}
where $w_0$ and $w_1$ are functions of $x$ and $t$. Note that in
the system \eqref{e1}, in the seventh term of the right-hand side
of its second equation, we have corrected a misprint made in
\cite{BDP}: the degree of $w_0$ should be $-7/2$ there.

In the present paper, we find chains of transformations which
relate these new generalized Harry Dym (GHD) equations \eqref{e1}
and \eqref{e2} with previously known integrable classical systems.
In Section~\ref{s2}, the GHD equation \eqref{e1} is transformed
into a pair of decoupled KdV equations. In Section~\ref{s3}, the
GHD equation \eqref{e2} is transformed into a pair of coupled mKdV
equations which belongs to the bi-Hamiltonian hierarchy of the
modified dispersive water waves equation of Kupershmidt \cite{Kup}
(see also \cite{Bla}, p.~84). Section~\ref{s4} contains concluding
remarks.

\section{Transforming the first GHD equation} \label{s2}

There are no general methods of transforming a given nonlinear
system into another one, less complicated or better studied. The
usual way of finding necessary transformations is based on
experience, guess and good luck. For this reason, we give no
comments on how these transformations were found in the present
case.

First, the transformation
\begin{equation}
\label{e3}
w_0 = u(x,t)^{-2}, \qquad w_1 = v(x,t), \qquad t \mapsto -4t
\end{equation}
brings the GHD equation \eqref{e1} into the following simpler
form:
\begin{equation}
\label{e4}
\begin{split}
u_t & = u^3 u_{xxx} , \\
v_t & = u^3 v_{xxx} + 9 u^2 v_x u_{xx}
+ 12 u^2 u_x v_{xx} + 27 u v u_x u_{xx} \\
& + \tfrac{57}{2} v u_x^3 + \tfrac{75}{2} u u_x^2 v_x
+ \tfrac{9}{8} u^6 v^3 u_x + \tfrac{3}{8} u^7 v^2 v_x .
\end{split}
\end{equation}

Second, we try to transform $x$, $u$ and $v$ in \eqref{e4}
as follows:
\begin{equation}
\label{e5}
x = p(y,t), \qquad u(x,t) = p_y(y,t), \qquad v(x,t) = q(y,t).
\end{equation}
This is an extension of the transformation used by Ibragimov
\cite{Ibr} to relate the original Harry Dym equation with the
Schwarzian-modified KdV equation. In the case of scalar evolution
equations, the Ibragimov transformation (i.e.\ \eqref{e5} with
$v=q=0$) is an essential link in chains of transformations between
constant separant equations and non-constant separant ones
\cite{S1,S2}. The transformation \eqref{e5} really works and
relates the system \eqref{e4} with the system
\begin{equation}
\label{e6}
\begin{split}
p_t & = p_{yyy} - \tfrac{3}{2} p_y^{-1} p_{yy}^2 ,\\
q_t & = q_{yyy} + 9 p_y^{-1} p_{yyy} q_y
+ 27 p_y^{-2} p_{yy} p_{yyy} q + 18 p_y^{-2} p_{yy}^2 q_y \\
& + 9 p_y^{-1} p_{yy} q_{yy} + \tfrac{3}{2} p_y^{-3} p_{yy}^3 q
+ \tfrac{3}{8} p_y^6 q^2 q_y + \tfrac{9}{8} p_y^5 p_{yy} q^3 .
\end{split}
\end{equation}
To verify this, one may use the following identities:
\begin{equation}
\label{e7}
u \partial_x = \partial_y, \qquad
u_t = p_{yt} - p_y^{-1} p_{yy} p_t, \qquad
v_t = q_t - p_y^{-1} q_y p_t.
\end{equation}
Note that \eqref{e5} is not an invertible transformation: it maps
the system \eqref{e6} into the system \eqref{e4}, whereas its
application in the opposite direction, from \eqref{e4} to
\eqref{e6}, requires one integration by $y$. We have omitted the
terms $\alpha (t) p_y$ and $\alpha (t) q_y$ in the right-hand
sides of the first and second equations of \eqref{e6},
respectively, where this arbitrary function $\alpha (t)$ appeared
as a `constant' of that integration.

Third, we make the transformation
\begin{equation}
\label{e8}
f(y,t) = p_y^{-1} p_{yy}, \qquad g(y,t) = p_y^3 q,
\end{equation}
admitted by the system \eqref{e6} owing to the form of its
equations, and obtain the pair of decoupled mKdV equations
\begin{equation}
\label{e9}
f_t = \left( f_{yy} - \tfrac{1}{2} f^3 \right)_y, \qquad
g_t = \left( g_{yy} + \tfrac{1}{8} g^3 \right)_y.
\end{equation}

Needless to say that the pair of Miura transformations
\begin{equation}
\label{e10}
a(y,t) = \pm f_y - \tfrac{1}{2} f^2, \qquad
b(y,t) = \pm \tfrac{1}{2} \mathrm{i} g_y + \tfrac{1}{8} g^2,
\end{equation}
with independent choice of the $\pm$ signs, relates \eqref{e9}
with the two copies of the KdV equation
\begin{equation}
\label{e11}
a_t = a_{yyy} + 3 a a_y, \qquad b_t = b_{yyy} + 3 b b_y.
\end{equation}

\section{Transforming the second GHD equation} \label{s3}

We follow the same three-step transformation as used in
Section~\ref{s2}.

First, the transformation \eqref{e3} brings the GHD equation
\eqref{e2} into the form
\begin{equation}
\label{e12}
\begin{split}
u_t & = u^3 u_{xxx} - \tfrac{9}{4} u^7 v^2 u_x
- \tfrac{3}{4} u^8 v v_x ,\\
v_t & = 3 u^2 v u_{xxx} + u^3 v_{xxx} + 36 u v u_x u_{xx}
+ 12 u^2 v_x u_{xx}\\
& + 12 u^2 u_x v_{xx} + 24 v u_x^3 + 36 u u_x^2 v_x
+ \tfrac{9}{4} u^6 v^3 u_x + \tfrac{3}{4} u^7 v^2 v_x .
\end{split}
\end{equation}

Second, we apply the transformation \eqref{e5} to the system
\eqref{e12} and obtain
\begin{equation}
\label{e13}
\begin{split}
p_t & = p_{yyy} - \tfrac{3}{2} p_y^{-1} p_{yy}^2
- \tfrac{3}{8} p_y^7 q^2 ,\\
q_t & = 3 p_y^{-1} p_{yyyy} q + 24 p_y^{-2} p_{yy} p_{yyy} q
+ 12 p_y^{-1} p_{yyy} q_y + q_{yyy} \\
& - 3 p_y^{-3} p_{yy}^3 q + \tfrac{27}{2} p_y^{-2} p_{yy}^2 q_y
+ 9 p_y^{-1} p_{yy} q_{yy} + \tfrac{9}{4} p_y^5 p_{yy} q^3
+ \tfrac{3}{8} p_y^6 q^2 q_y ,
\end{split}
\end{equation}
where the terms $\alpha (t) p_y$ and $\alpha (t) q_y$, with
arbitrary $\alpha (t)$, have been omitted in the right-hand
sides of the first and second equations, respectively.

Third, the transformation \eqref{e8} relates the system
\eqref{e13} with the following system of coupled mKdV
equations:
\begin{equation}
\label{e14}
\begin{split}
f_t & = \left( f_{yy} - \tfrac{3}{4} g g_y
- \tfrac{1}{2} f^3 - \tfrac{3}{8} f g^2 \right)_y , \\
g_t & = \left( g_{yy} + 3 g f_y
- \tfrac{3}{2} f^2 g - \tfrac{5}{8} g^3 \right)_y .
\end{split}
\end{equation}

The system \eqref{e14} does not admit any further
transformation into a system of coupled KdV equations. It is
possible to transform \eqref{e14} into a system of a
KdV--mKdV type, but we will not follow this way. Instead, we
notice that the system \eqref{e14} is invariant under the
change of variables $f \mapsto f$, $g \mapsto - g$. Therefore
the transformation
\begin{equation}
\label{e15}
f = c_1 ( a + b ) , \qquad g = c_2 ( a - b ) ,
\end{equation}
with any nonzero constants $c_1$ and $c_2$, relates the system
\eqref{e14} with a system of symmetrically coupled mKdV equations
for $a(y,t)$ and $b(y,t)$, which is invariant under $a \mapsto b$,
$b \mapsto a$. Systems of symmetrically coupled mKdV equations
possessing higher-order generalized symmetries were classified by
Foursov \cite{Fou}. The choice of
\begin{equation}
\label{e16}
c_1 = 1, \qquad c_2 = \pm \mathrm{i}
\end{equation}
in the transformation \eqref{e15} brings the system \eqref{e14}
into the form
\begin{equation}
\label{e17}
\begin{split}
a_t & = \left( a_{yy} + 3 a a_y - 3 b a_y + a^3
- 6 a^2 b + 3 a b^2 \right)_y , \\
b_t & = \left( b_{yy} + 3 b b_y - 3 a b_y + b^3
- 6 b^2 a + 3 b a^2 \right)_y ,
\end{split}
\end{equation}
which is exactly the case (K) in the Foursov classification
\cite{Fou}.

Foursov \cite{Fou} proved that the system \eqref{e17} represents
the third-order generalized symmetry of the system of coupled
Burgers equations
\begin{equation}
\label{e18}
a_t = \left( a_y + a^2 - 2 a b \right)_y , \qquad
b_t = \left( - b_y + 2 a b - b^2 \right)_y ,
\end{equation}
and found the bi-Hamiltonian structure of this hierarchy with
the Hamiltonian operators
\begin{equation}
\label{e19}
\begin{split}
P & = \begin{pmatrix} 0 & \partial_y \\
\partial_y & 0 \end{pmatrix} , \\
Q & = \begin{pmatrix} - 2 a \partial_y - a_y
& \partial_y^2 + ( a - b ) \partial_y + a_y \\
- \partial_y^2 + ( a - b ) \partial_y - b_y
& 2 b \partial_y + b_y \end{pmatrix} .
\end{split}
\end{equation}

In its turn, the system of coupled Burgers equations \eqref{e18}
has a long history. As a system of coupled second-order evolution
equations possessing higher-order symmetries, it appeared in the
classifications of Mikhailov, Shabat and Yamilov \cite{MSY} and
Olver and Sokolov \cite{OS}. Moreover, the bi-Hamiltonian
structure \eqref{e19} turns out to be not new. Indeed, the
transformation
\begin{equation}
\label{e20}
a = - r , \qquad b = s - r , \qquad t \mapsto - \tfrac{1}{2} t
\end{equation}
relates the system \eqref{e18} with the modified dispersive
water waves equation
\begin{equation}
\label{e21}
\begin{split}
r_t & = \tfrac{1}{2} \left( - r_y + 2 r s - r^2 \right)_y , \\
s_t & = \tfrac{1}{2} \left( s_y - 2 r_y - 2 r^2 + 2 r s + s^2
\right)_y
\end{split}
\end{equation}
which was introduced, together with its bi-Hamiltonian structure,
by Kupershmidt \cite{Kup} (see also \cite{Bla}, p.~84). The
bi-Hamiltonian structures of \eqref{e18} and \eqref{e21} are
related by the transformation \eqref{e20} as well. For this
reason, the system \eqref{e17} is equivalent to a third-order
member of the bi-Hamiltonian hierarchy of the modified dispersive water
waves equation \eqref{e21}.

\section{Conclusion} \label{s4}

In this paper, we found chains of transformations which relate
the new GHD equations \eqref{e1} and \eqref{e2} of Brunelli, Das
and Popowicz with previously known integrable systems. The
transformations \eqref{e3}, \eqref{e5}, \eqref{e8} and
\eqref{e10} relate the GHD equation \eqref{e1} with the pair of
decoupled KdV equations \eqref{e11}. The transformations
\eqref{e3}, \eqref{e5}, \eqref{e8}, \eqref{e15} with the choice
of \eqref{e16}, and \eqref{e20} relate the GHD equation
\eqref{e2} with a third-order member of the bi-Hamiltonian
hierarchy of the modified dispersive water waves equation
\eqref{e21}.

It can be observed in the literature (see, e.g.,
\cite{Ibr,S1,S2,MSY} and references therein) that quite often a
newly-found remarkable equation turns out to be related to a
well-studied old equation through an explicit chain of
transformations. In such a situation, one gets a possibility not
to study the new equation directly but to derive its properties
from the well-known properties of the corresponding old equation,
using the transformations obtained. Now this applies to the new
generalized Harry Dym equations of Brunelli, Das and Popowicz as
well.

\section*{Acknowledgments}

The author is grateful to Dr.~A.~Sergyeyev for some copies of
references, to the Mathematical Institute of Silesian University
for hospitality, and to the Centre for Higher Education Studies
of Czech Republic for support.

\end{document}